\documentclass[12pt]{article}
\usepackage{amsthm,amsmath,indentfirst,amssymb}
\usepackage{algorithm,algpseudocode}

\usepackage{amsmath,amsthm,amsfonts,amssymb,indentfirst}
\usepackage{algorithm, algpseudocode}
\usepackage{graphicx}
\usepackage{subcaption}
\usepackage{color}
\newtheorem{thm}{Theorem}[section]
\newtheorem{cor}[thm]{Corollary}
\newtheorem{lem}[thm]{Lemma}
\newtheorem{op}[thm]{Problem}

\newtheorem{pro}[thm]{Proposition}

\newtheorem{obs}[thm]{Observation}

\newenvironment{pf}{{\noindent \it \bf Proof:}}{{\hfill$\Box$}\\}
\begin{document}

\title{\bf Extremality and Sharp Bounds for the $k$-edge-connectivity of Graphs}

\author{Yuefang Sun$^{1}$\thanks{E-mail: yuefangsun2013@163.com. This author was supported by National Natural Science Foundation of China
(No.11401389), China Scholarship Council (No.201608330111) and
Zhejiang Provincial Natural Science Foundation of China
(No.LY17A010017).}\quad Xiaoyan Zhang$^{2}$\thanks{E-mail:
royxyzhang@gmail.com.  This author was supported by National Natural Science Foundation of China
(No.11471003 and No.11871280) and Qing Lan Project. (Corresponding author.)} \quad Zhao Zhang$^{3}$\thanks{E-mail:
hxhzz@sina.com. This author was supported by National Natural Science Foundation of China
(No.11531011 and No.11771013).}\\ \\
$^{1}$ Department of Mathematics, Shaoxing University\\ Zhejiang 312000, China\\
$^{2}$ School of Mathematical Science \& Institute of Mathematics\\ Nanjing Normal University,\\ Jiangsu 210023, China\\
$^{3}$ College of Mathematics and Computer Science \\Zhejiang Normal University,\\Zhejiang 321004, China\\
}

\date{}
\maketitle

\begin{abstract}
Boesch and Chen (SIAM J. Appl. Math., 1978) introduced
the cut-version of the generalized edge-connectivity, named
$k$-edge-connectivity. For any integer $k$ with $2\leq k\leq n$, the
{\em $k$-edge-connectivity} of a graph $G$, denoted by
$\lambda_k(G)$, is defined as the smallest number of edges whose
removal from $G$ produces a graph with at least $k$ components.

In this paper, we first compute some exact values and sharp bounds
for $\lambda_k(G)$ in terms of $n$ and $k$. We then discuss the
relationships between $\lambda_k(G)$ and other generalized
connectivities. An algorithm in $\mathcal{O}(n^2)$ time will be
provided such that we can get a sharp upper bound in terms of the
maximum degree. Among our results, we also compute some exact values
and sharp bounds for the function $f(n,k,t)$ which is defined as the
minimum size of a connected graph $G$ with order $n$ and
$\lambda_k(G)=t$. \\
[2mm]{\bf Key Words}: $k$-edge-connectivity, $k$-connectivity, generalized
connectivity, $S$-Steiner tree, $k$-way cut. \\
\end{abstract}


\section{Introduction}

We refer to \cite{Bondy} for graph theoretical notation and
terminology not described here. For a graph $G$, let $V(G)$, $E(G)$
be the set of vertices, the set of edges of $G$, respectively. For
$X\subseteq V(G)$, we denote by $G-X$ the subgraph obtained by
deleting from $G$ the vertices of $X$ together with the edges
incident with them. For $Y\subseteq E(G)$, we denote by $G-Y$ the
subgraph obtained by deleting from $G$ the edges of $Y$. For a set
$S$, we use $|S|$ to denote its size. We use $P_n$, $C_m$ and
$K_{\ell}$ to denote a path of order $n$, a cycle of order $m$ and a
complete graph of order $\ell$, respectively.

Connectivity is one of the most basic concepts in graph theory, both
in combinatorial sense and in algorithmic sense, see\cite{Bondy,
Chrobak-Karloff-Radzik,Godsil-Royle,Provan-Ball}.
The edge-connectivity of $G$, written by $\lambda(G)$, is the
minimum size of an edge set $Y\subseteq E(G)$ such that $G-Y$ is
disconnected. This definition is called the {\em cut-version}
definition of the edge-connectivity. A well-known theorem of Menger
provides an equivalent definition, which can be called the {\em
path-version} definition of the edge-connectivity. For any two
distinct vertices $x$ and $y$ in $G$, the local edge-connectivity
$\lambda_G(x,y)$ is the maximum number of edge-disjoint paths
connecting $x$ and $y$. Then $\lambda(G) = \min\{\lambda_G(x,y)\mid
x,y \in V(G), x\neq y\}$ is defined to be the edge-connectivity of
$G$. Similarly, there are cut-version and path-version definitions
for the connectivity of graphs.

In \cite{Boesch-Chen}, Boesch and Chen generalized the cut-version
definition of the edge-connectivity, which has many applications in practice. However, they did not give a
specific name for such a generalized edge-connectivity. Here we will
use the name ``$k$-edge-connectivity'' from \cite{Li-Mao5}. For any
integer $k$ with $2\leq k\leq n$, the {\em $k$-edge-connectivity} of
a graph $G$, denoted by $\lambda_k(G)$, is defined as the smallest
number of edges whose removal from $G$ produces a graph with at
least $k$ components. By definition, we clearly have
$\lambda_2(G)=\lambda(G)$. Boesch and Chen \cite{Boesch-Chen} gave
some properties of $\lambda_k(G)$ and obtained some bounds for
$\lambda_k(G)$ in terms the minimum degree and the degree-sequence
of $G$. They also studied some special cases, such as complete
graphs.

The problem of $k$-edge-connectivity is also called the {\em $k$-WAY
CUT problem} which is defined as follows: given an undirected graph
$G$ and integers $k$ and $s$, remove at most $s$ edges from $G$ to
obtain a graph with at least $k$ connected components. This problem
has applications in numerous areas of computer science, such as
finding cutting planes for the traveling salesman problem,
clustering-related settings (e.g., VLSI design), or network
reliability \cite{Burlet-Goldschmidt}. In \cite{Cygan, Downey,
Goldschmidt-Hochbaum, Kamidoi-Yoshida-Nagamochi,
Kawarabayashi-Thorup}, the authors considered the algorithms and
computational complexity of this problem. In general, $k$-WAY CUT is
NP-complete \cite{Goldschmidt-Hochbaum} but solvable in polynomial
time for fixed $k$ \cite{Goldschmidt-Hochbaum,
Kamidoi-Yoshida-Nagamochi}. From the parameterized perspective, the
$k$-WAY CUT problem parameterized by $k$ is W[1]-hard \cite{Downey}.
Kawarabayashi and Thorup \cite{Kawarabayashi-Thorup} presented a
fixed-parameter algorithm for $k$-WAY CUT parameterized by $s$. In
\cite{Cygan}, Cygan et al. showed that it is OR-compositional and,
therefore, a polynomial kernelization algorithm is unlikely to
exist.

In this paper, we continue to compute the exact values and sharp bounds of
$k$-edge-connectivity for a graph $G$, and investigate the extremality for the $\lambda_k(G)$ of graphs. Some concepts and preliminary results will be introduced in the next section. In
Section 3, we will characterize those graphs $G$ with
$\lambda_k(G)=t$, where $t\in \{k-1, {n\choose 2}-{{n-k+1}\choose
2}-1, {n\choose 2}-{{n-k+1}\choose 2}\}$. For any connected graph
$G$, we will obtain a sharp lower and a sharp upper bounds of
$\lambda_k(G)$ in terms of $n$ and $k$, and we will give necessary
and sufficient conditions for equalities to hold.

Relationships between $\lambda_k(G)$ and other generalized
connectivities, including $\lambda'_k(G), \kappa'_k(G)$ and
$\kappa_k(G)$, will also be discussed in Section 3. Note that
definitions of these generalized connectivities will be introduced
in Section 2. We will first compute a sharp lower bound which is about
the relationship between $\lambda_k(G)$ and $\lambda'_k(G)$
($\kappa'_k(G)$), and a sharp upper bound which concerns the
relationship between $\lambda_k(G)$ and $\kappa_{k-1}(G)$. Moreover,
a sharp bound that $\lambda_k(G)\geq \kappa_{k}(L(G))$ will also be
deduced, where $L(G)$ is the line graph of $G$.

An algorithm in $\mathcal{O}(n^2)$ time will be provided such that
we can compute a sharp upper bound in terms of the maximum degree
$\Delta(G)$ of a graph $G$.

For $2\leq k\leq n$ and $k-1\leq t\leq {n\choose 2}-{{n-k+1}\choose
2}$, the function $f(n,k,t)$ is defined as the minimum size of a
connected graph $G$ with order $n$ and $\lambda_k(G)=t$. Bounds and
some exact values for this function will be computed.

\section{Preliminaries}

We now introduce concepts of three generalized (edge-)
connectivities which will be useful in our argument. Chartrand et
al. \cite{Chartrand1} generalized the cut-version definition of the
connectivity as follows: For an integer $k\geq 2$ and a graph $G$ of
order $n\geq k$, the {\em $k$-connectivity} $\kappa_k(G)$ is the
smallest number of vertices whose removal from $G$ produces a graph
with at least $k$ components or a graph with fewer than $k$
vertices. By definition, we clearly have $\kappa_2(G)=\kappa(G)$.
For more details about this topic, we refer to \cite{Chartrand1,
Oellermann1, sun9, sun6}.

The generalized $k$-connectivity $\kappa'_k(G)$ of a graph $G$ which
was introduced by Hager \cite{Hager} in 1985 is a natural
generalization of the path-version definition of the connectivity.
For a graph $G=(V,E)$ and a set $S\subseteq V$ of at least two
vertices, an {\em $S$-Steiner tree} or a {\em Steiner tree
connecting $S$} (or simply, an {\em $S$-tree}) is a such subgraph
$T$ of $G$ that is a tree with $S\subseteq V(T)$. Two $S$-trees
$T_1$ and $T_2$ are said to be {\em internally disjoint} if
$E(T_1)\cap E(T_2)=\emptyset$ and $V(T_1)\cap V(T_2)=S$. The {\em
generalized local connectivity} $\kappa'_G(S)$ is the maximum number
of internally disjoint $S$-trees in $G$. For an integer $k$ with
$2\leq k\leq n$, the {\em generalized $k$-connectivity} is defined
as $$\kappa'_k(G)=\min\{\kappa'_G(S)\mid S\subseteq V(G), |S|=k\}.$$
Thus, $\kappa'_k(G)$ is the minimum value of $\kappa'_G(S)$ when $S$
runs over all the $k$-subsets of $V(G)$. By definition, we clearly
have $\kappa'_2(G)=\kappa(G)$. By convention, for a connected graph
$G$ with less than $k$ vertices, we set $\kappa'_k(G) = 1$, and
$\kappa'_k(G) = 0$ when $G$ is disconnected. For more details about
this topic, the reader can see \cite{Hager, H.Li, Li-Mao-Sun, Li-Mao3, sun6}.

As a natural counterpart of the generalized $k$-connectivity, Li,
Mao and Sun \cite{Li-Mao-Sun} introduced the following concept of
generalized edge-connectivity which is a generalization of the
path-version definition of the edge-connectivity. Two $S$-trees
$T_1$ and $T_2$ are said to be {\em edge-disjoint} if $E(T_1)\cap
E(T_2)=\emptyset$. The {\em generalized local edge-connectivity}
$\lambda'_G(S)$ is the maximum number of edge-disjoint $S$-trees in
$G$. For an integer $k$ with $2\leq k\leq n$, the {\em generalized
$k$-edge-connectivity} is defined as
$$\lambda'_k(G)=\min\{\lambda'_G(S)\mid S\subseteq V(G), |S|=k\}.$$ Thus,
$\lambda'_k(G)$ is the minimum value of $\lambda'_G(S)$ when $S$
runs over all the $k$-subsets of $V(G)$. Hence, we have
$\lambda'_2(G)=\lambda(G)$. By definitions of $\kappa'_k(G)$ and
$\lambda'_k(G)$, $\kappa'_k(G)\leq \lambda'_k(G)$ holds. By
definitions, the generalized local edge-connectivity is the famous
Steiner Packing Problem, see \cite{Frank, Lau, West2}.

Nowadays, more and more researchers are working in the topic of
generalized connectivity with applications. There are many results
on this type of generalized edge-connectivity, such as
\cite{Li-Mao4, Li-Mao-Sun}. The reader is also referred to a new
book \cite{Li-Mao5} for a detailed introduction of this field.

The following two observations can be obtained straightforwardly
from the definition of $\lambda_k(G)$.

\begin{obs}\label{thm1}
Let $H$ be a connected spanning subgraph of a graph $G$, we have
$\lambda_k(H)\leq \lambda_k(G)$.
\end{obs}

\begin{obs}\label{thm2}
For any integer $k$ with $2\leq k\leq n-1$, we have
$\lambda_k(G)\leq \lambda_{k+1}(G)$.
\end{obs}

In the rest of this section, we will present exact values of
$\lambda_k(G)$ for some special graph classes which will be used in
our argument of the main results for general graphs given in the next section. A {\em wheel graph} $W_n$ of order $n$ is a graph that
contains a cycle of order $n-1$, and every graph vertex in the cycle
is connected to one other graph vertex, which is known as the {\em
hub}.

\begin{lem}\label{thm5}
The following assertions hold:\\
$(i)$\cite{Boesch-Chen}~$\lambda_k(T)= k-1$, where $T$ is a tree;\\
$(ii)$~$\lambda_k(C_n)=k$;\\
$(iii)$~$\lambda_k(W_n)= \left\{
\begin{array}{ll}
2k-1, & if~2\leq k\leq n-1\\
2k-2, & if~k= n.
\end{array}
\right.$
\end{lem}
\begin{pf} The assertion$(i)$ is from \cite{Boesch-Chen}. The assertion
$(ii)$ is not hard, so we omit the details. We now prove $(iii)$ and
assume that $2\leq k\leq n-1$ in the following argument since the
special case that $k=n$ is clear. Let $G\cong W_n$ such that $C:
v_1, v_2, \cdots, v_{n-1}$ is the cycle of order $n-1$ and $u$ is
the hub. Let $Y_0$ be the set of edges incident to elements of
$\{v_i\mid 1\leq i\leq k-1\}$. Clearly, $|Y_0|=2k-1$ and $G-Y_0$
contains $k$ components, then $\lambda_k(W_n)\leq 2k-1$.

Let $Y\subseteq E(G)$ with $|Y|\leq 2k-2$. Suppose that $G'=G-Y$
contains $\ell$ components: $G_1, G_2, \cdots, G_{\ell}$, where
$\ell \geq k$. Without loss of generality, we assume that $u\in
V(G_1)$ and will consider the following two cases:

{\bf Case 1.} $V(G_1)=\{u\}$. In this case all edges incident to $u$
must belong to $Y$ and then $Y \cap E(C)\leq (2k-2)-(n-1)\leq k-2$,
so $G'[V(C)]= C-(Y \cap E(C))$ contains at most $k-2$ components.
Since $G'=G-Y$ contains at least $k$ components and $G_1$ is a
trivial one with $\{u\} = V(G_1)$, we have that $G'[V(C)]$ has at
least $k-1$ components, a contradiction.

{\bf Case 2.} $V(G_1)\setminus \{u\}\neq \emptyset$. Let $E_u$ be
the set of edges incident to $u$ in $G$. We know that in $G'$ there
is no edge between $u$ and $V(G_i)$ for $2\leq i\leq \ell$. Then $|Y
\cap E_u| \geq \ell-1\geq k-1$ and so $|Y \cap E(C)|\leq
(2k-2)-(k-1)=k-1$. Hence, $G'[V(C)]= C-(Y \cap E(C))$ contains at
most $k-1$ components. Since $G'=G-Y$ contains at least $k$
components and $V(G_1)\setminus \{u\}\neq \emptyset$, we know that
$G'[V(C)]$ has at least $k$ components, a contradiction.

By the above argument, we have $\lambda_k(W_n)\geq 2k-1$, and
furthermore $\lambda_k(W_n)= 2k-1$.
\end{pf}

\begin{lem}\label{thm01}\cite{sun9} Let $G$ be a graph with order $n$ and size $m$.
If $G$ contains at least $k$ components, then $m \leq
{{n-k+1}\choose 2}$; the equality holds if and only if $G$ has
exactly $k$ components such that $k-1$ of them are trivial, the
remaining one is a clique of order $n -k + 1$.
\end{lem}

Note that Boesch and Chen \cite{Boesch-Chen} have determined the
precise value for $\lambda_k(K_n)$. Here, we restate their result
with a different argument which will be useful in the following
discussion.

\begin{lem}\label{thm3}\cite{Boesch-Chen}
$\lambda_k(K_n)= {n\choose 2}-{{n-k+1}\choose 2}.$
\end{lem}
\begin{pf}
Let $G\cong K_n$ with vertex set $V(G)=\{u_i\mid 1\leq i\leq n\}$.
Let $Y_0$ be a set of edges which are incident to any member of
$\{u_i\mid 1\leq i\leq k-1\}$. Clearly, $|Y_0|={n\choose
2}-{{n-k+1}\choose 2}$, and the graph $G-Y_0$ contains exactly $k$
components: a clique with order $n-k+1$ and $k-1$ trivial
components. Hence, $\lambda_k(K_n)\leq |Y_0|={n\choose
2}-{{n-k+1}\choose 2}.$

Let $Y$ be any edge set of $G$ such that $G-Y$ contains $\ell$
components, $G_1, G_2, \cdots, G_{\ell}$, where $\ell \geq k$.
Without loss of generality, we can assume that
$n_1\leq n_2\leq \cdots \leq n_{\ell}$, where $n_i=n(G_i)$. 
Since $\sum_{i=1}^k{n_i}=n$ and $1\leq n_i \leq n-k+1$, by Lemma
\ref{thm01}, it is not hard to show that $\sum_{i=1}^{\ell}{E(G_i)}$
attains the maximum value if and only if $\ell=k,
n_1=n_2=\cdots=n_{k-1}=1, n_k=n-k+1$, and each $G_i$ is a clique,
that is, $\sum_{i=1}^{\ell}{E(G_i)}\leq
\frac{k-1}{2}+\frac{(n-k+1)^2}{2}-\frac{n}{2}={{n-k+1}\choose 2}$
with the equality holds only if $n_1=n_2=\cdots=n_{k-1}=1,
n_k=n-k+1$, and each $G_i$ is a clique. Then $|Y|= {n\choose 2}-
\sum_{i=1}^k{E(G_i)}\geq {n\choose 2}-{{n-k+1}\choose 2}$ and so
$\lambda_k(K_n)\geq {n\choose 2}-{{n-k+1}\choose 2}.$ This completes
the proof.
\end{pf}

We use $K_n-e$ to denote a graph obtained from a complete graph
$K_n$ by deleting any edge $e$. Then, we have the following lemma.
\begin{lem}\label{thm4}
$\lambda_k(K_n-e)= {n\choose 2}-{{n-k+1}\choose 2}-1.$
\end{lem}
\begin{pf}
Let $G\cong K_n-e$ with vertex set $V(G)=\{u_i\mid 1\leq i\leq n\}$
such that $u_1v_1\not\in E(G)$. Let $Y_0$ be a set of edges which
are incident to any member of $\{u_i\mid 1\leq i\leq k-1\}$.
Clearly, $|Y_0|={n\choose 2}-{{n-k+1}\choose 2}-1$, and the graph
$G-Y_0$ contains exactly $k$ components: a clique with order $n-k+1$
and $k-1$ trivial components. Hence, $\lambda_k(K_n)\leq
|Y_0|={n\choose 2}-{{n-k+1}\choose 2}-1.$

Let $Y$ be any edge set of $G$ such that $G-Y$ contains $\ell$
components, $G_1, G_2, \cdots, G_{\ell}$, where $\ell \geq k$. With
a similar argument to that of Lemma \ref{thm3}, we have
$\sum_{i=1}^{\ell}{E(G_i)}\leq {{n-k+1}\choose 2}$. Then $|Y|=
{n\choose 2}-1- \sum_{i=1}^{\ell}{E(G_i)}\geq {n\choose
2}-{{n-k+1}\choose 2}-1$ and so $\lambda_k(K_n)\geq {n\choose
2}-{{n-k+1}\choose 2}-1.$ This completes the proof.
\end{pf}

We still need the following lemma.

\begin{lem}\label{thm6}If $G\not\cong K_n, K_n-e$, then
$\lambda_k(G)\leq {n\choose 2}-{{n-k+1}\choose 2}-2.$
\end{lem}
\begin{pf} Let $G$ be a graph obtained from a complete graph $K_n$
by deleting two edges $e_1,e_2$. Let $V(G)=\{u_i\mid 1\leq i\leq
n\}$. In the following, we will show that $\lambda_k(G)\leq
{n\choose 2}-{{n-k+1}\choose 2}-2$, and then our result clearly
holds by Observation \ref{thm1}. We will consider two cases
according to whether $e_1$ and $e_2$ are adjacent.

{\bf Case 1.} $e_1$ and $e_2$ are adjacent. Without loss of
generality, we assume that $e_1=u_1u_2$ and $e_2=u_1u_3$. Let $Y_0$
be a set of edges which are incident to any member of $\{u_i\mid
1\leq i\leq k-1\}$. Clearly, $|Y_0|={n\choose 2}-{{n-k+1}\choose
2}-2$, and the graph $G-Y_0$ contains exactly $k$ components: a
clique with order $n-k+1$ and $k-1$ trivial components. Hence,
$\lambda_k(K_n)\leq |Y_0|={n\choose 2}-{{n-k+1}\choose 2}-2.$

{\bf Case 2.} $e_1$ and $e_2$ are nonadjacent. Without loss of
generality, we assume that $e_1=u_1u_2$ and $e_2=u_3u_4$.

We first consider the case that $k\geq 4$. Let $Y_1$ be a set of
edges which are incident to any member of $\{u_i\mid 1\leq i\leq
k-1\}$. Clearly, $|Y_1|={n\choose 2}-{{n-k+1}\choose 2}-2$, and the
graph $G-Y_1$ contains exactly $k$ components: a clique with order
$n-k+1$ and $k-1$ trivial components. Hence, $\lambda_k(K_n)\leq
|Y_1|={n\choose 2}-{{n-k+1}\choose 2}-2.$

We then consider the case that $k=3$. Let
$Y_2=\{u_1u_4,u_1u_3,u_2u_4\}\cup E(\{u_1,u_4\}, A)$, where
$A=V\setminus \{u_i\mid 1\leq i\leq 4\}$ and $E(\{u_1,u_4\}, A)$
denotes the set of edges between $\{u_1,u_4\}$ and $A$. Clearly,
$|Y_2|=2n-5={n\choose 2}-{{n-k+1}\choose 2}-2$, and the graph
$G-Y_2$ contains exactly three components: a clique with order $n-2$
and two trivial components. Hence, $\lambda_k(K_n)\leq
|Y_2|={n\choose 2}-{{n-k+1}\choose 2}-2.$
\end{pf}


\section{Main results of computing exact values and sharp bounds}

By Lemmas \ref{thm3}, \ref{thm4} and \ref{thm6}, the following
result clearly holds.

\begin{pro}\label{thm7}The following assertions hold:\\
$(i)$~$\lambda_k(G)= {n\choose 2}-{{n-k+1}\choose 2}-1$ if and only
if $G\cong K_n-e$; \\
$(ii)$~$\lambda_k(G)= {n\choose 2}-{{n-k+1}\choose 2}$ if and only
if $G\cong K_n$.
\end{pro}

The following result concerns sharp bounds for $\lambda_k(G)$ of a
general graph $G$.
\begin{thm}\label{thma}
For a connected graph $G$, we have $$k-1\leq \lambda_k(G)\leq
{n\choose 2}-{{n-k+1}\choose 2}.$$ Moreover, the lower bound can be
attained if and only if $G$ contains at least $k-1$ cut edges, and
the upper bound can be attained if and only if $G\cong K_n$.
\end{thm}
\begin{pf} The lower bound is clear by Observation \ref{thm1} and $(i)$
of Lemma \ref{thm5}. If $G$ contains at least $k-1$ cut edges, then
let $Y_0$ be a set of $k-1$ cut edges. Clearly, $G-Y_0$ contains $k$
components and so $\lambda_k(G)\leq k-1$. Hence, $\lambda_k(G)= k-1$
in this case. If $G$ contains at most $k-2$ cut edges, then let
$Y=\{e_1, e_2, \cdots, e_{k-1}\}$ be a set of any $k-1$ edges of
$G$. Without loss of generality, we assume that the former $k_1$
elements of $Y$ are cut edges. Then $G'=G-\{e_i\mid 1\leq i\leq
k_1\}$ has exactly $k_1+1$ components. We know that each element of
$\{e_i|k_1+1\leq i\leq k-1\}$ is not a cut edge of $G'$ and so the
number of components in $G''=G-Y=G'-\{e_i\mid 1\leq i\leq k_1\}$
will increase at most $(k-1-k_1)-1$. Hence, the number of components
in $G''$ is at most $(k-1-k_1)-1+k_1+1=k-1$ components and so
$\lambda_k(G)> |Y|=k-1$, a contradiction. Therefore, the lower bound
can be attained if and only if $G$ contains at least $k-1$ cut
edges.

We now prove the upper bound. By Observation \ref{thm1} and Lemma
\ref{thm3}, we have $\lambda_k(G)\leq \lambda_k(K_n)= {n\choose
2}-{{n-k+1}\choose 2}.$ By Proposition \ref{thm7}, the upper bound
can be attained if and only if $G\cong K_n$.
\end{pf}

Note that by Proposition \ref{thm7} and Theorem \ref{thma}, we can
characterize those graphs $G$ with $\lambda_k(G)=t$ for $t\in \{k-1,
{n\choose 2}-{{n-k+1}\choose 2}-1, {n\choose 2}-{{n-k+1}\choose
2}\}$.

We now discuss the relationships between $\lambda_k(G)$ and other
generalized connectivities, including $\lambda'_k(G), \kappa'_k(G)$
and $\kappa_k(G)$. We first give a lower bound which concerns the
relationship between $\lambda_k(G)$ and $\lambda'_k(G)$, and an
upper bound which is about the relationship between $\lambda_k(G)$
and $\kappa_{k-1}(G)$.

\begin{thm}\label{thmc}
For a connected graph $G$ with maximum degree $\Delta(G)$, we have
$$(k-1){\lambda'_k(G)}\leq \lambda_k(G)\leq {\Delta(G)}{\kappa_{k-1}(G)}.$$ Moreover, both bounds
are sharp.
\end{thm}
\begin{pf}
We first prove the lower bound and its sharpness. For the case
$k=2$, the result clearly holds. In the following argument, we
assume that $k\geq 3$. Let $Y$ be a set of edges of $G$ with
$|Y|=\lambda_k(G)$ such that the graph $G-Y$ contains $\ell$
components, say $G_1, G_2, \cdots, G_{\ell}$, where $\ell\geq k$.
Let $S=\{u_i\mid u_i\in V(G_i), 1\leq i\leq k\}$. By the definition
of $\lambda'_k(G)$, there are at least $\lambda'_k(G)$ edge-disjoint
$S$-trees in $G$. For each such tree $T$, we have that $|E(T)\cap
Y|\geq k-1$, then $\lambda_k(G)=|Y|\geq (k-1){\lambda'_k(G)}.$

For the sharpness of the lower bound, we just consider the case that
$G$ is a tree. In this case we have that $\lambda'_k(G)=1$ and
$\lambda_k(G)=k-1$, so $\lambda_k(G)= (k-1){\lambda'_k(G)}.$

We now prove the upper bound and its sharpness. Let $X\subseteq
V(G)$ such that $|X|=\kappa_{k-1}(G)$ and $G-X$ contains at least
$k-1$ components. Let $E'$ be the set of edges between $X$ and
$V(G)\setminus X$ in $G$. Clearly, the graph $G-E'$ contains at
least $k$ components. Then $\lambda_k(G)\leq |E'|\leq
{\Delta(G)}{|X|}={\Delta(G)}{\kappa_{k-1}(G)}.$

For the sharpness of the upper bound, we just consider the following
graph $G$: Let $G$ be obtained by identifying the center vertex, say
$u_{k-1}$, of a star graph $S$ with an end vertex, say $v_{n-k+2}$,
of a path $P$ such that $V(S)=\{u_i\mid 1\leq i\leq k-1\}$,
$V(P)=\{v_j\mid 1\leq j\leq n-k+2\}$ and the new vertex of $G$ is
denoted by $u$. Clearly, $G$ is a tree with maximum degree
$\Delta(G)=deg_G(u)=k-1$ and so $\lambda_k(G)=k-1$. It is not hard
to show that $\kappa_{k-1}(G)=1$. Hence,
$\lambda_k(G)={\Delta(G)}{\kappa_{k-1}(G)}$ in this case.
\end{pf}

Note that for some graphs, the equalities of bounds in Theorem
\ref{thmc} may not hold. For the upper bound, let $G$ be a cycle
with order $n\geq 2k$, where $k\geq 3$. In this case we have that
$\kappa_{k-1}(G)=k-1$ and $\lambda_k(G)=k$, so $\lambda_k(G)<
{\Delta(G)}{\kappa_{k-1}(G)}$. For the lower bound, let $G$ be a
wheel graph with order $n\geq k+1$. It is not hard to show that
$\lambda'_k(G)\leq 2$. By Lemma \ref{thm5}, we have
$\lambda'_k(G)=2k-1$, so $(k-1){\lambda'_k(G)}<\lambda_k(G).$

Recall the fact that $\kappa'_k(G)\leq \lambda'_k(G)$ and by Theorem
\ref{thmc}, we have the following corollary. For the sharpness of
this bound, we just let $G$ be a tree.

\begin{cor}\label{thm9}
For a connected graph $G$, we have
$$\lambda_k(G)\geq (k-1){\kappa'_k(G)}.$$ Moreover, the bound
is sharp.
\end{cor}

The $line~graph$ $L(G)$ of a graph $G$ is the graph whose vertex set
is $V(L(G))=E(G)$ and two vertices $e_1$, $e_2$ of $L(G)$ are
adjacent if and only if these edges are adjacent in $G$. By using
the particular properties of line graphs shown in \cite{Hedetniemi}
and \cite{Hemminger}, we can give the following lower bound for
$\lambda_k(G)$ in terms of $k$-connectivity of the line graph of
$G$.

\begin{thm}\label{thme}
Let $G$ be a connected graph with order $n$. For $2\leq k\leq n$, we
have
$$\lambda_k(G)\geq \kappa_k(L(G)).$$
Moreover, the bound is sharp.
\end{thm}
\begin{pf} Let $Y=\{e_1, e_2, \cdots, e_t\}\subseteq E(G)$ with
$t=\lambda_k(G)$ such that $G-Y$ contains the following $k$
components: $G_1, G_2, \cdots, G_k$. Let $X=\{v_{e_1}, v_{e_2},
\cdots, v_{e_t}\}\subseteq V(L(G))$, where $v_{e_i}$ denotes the
vertex in $L(G)$ corresponding to the edge $e_i$ in $G$ for $1\leq
i\leq t$. It is not hard to show that ${L(G)}-X$ contains the
following $k$ components: $L(G_1), L(G_2), \cdots, L(G_k)$, where
$L(G_j)$ denotes the line graph of $G_j$ for $1\leq j\leq k$. Hence,
$\kappa_k(L(G))\leq |X|=|Y|=t=\lambda_k(G)$.

For the sharpness of the bound, we just let $G=C_n$ with $n\geq 2k$,
then we have $\lambda_k(G)=k$. Since $L(G)\cong C_n$, we obtain
$\kappa_k(L(G))=k$. Hence, $\lambda_k(G)= \kappa_k(L(G))$ in this
case.
\end{pf}

The following result is a sharp upper bound for $\lambda_k(G)$ which
is also a function of the maximum degree $\Delta(G)$.

\begin{thm}\label{thmd}
Let $G$ be a connected graph with order $n$ and maximum degree
$\Delta(G)$. For $2\leq k\leq n$, we have
$$\lambda_k(G)\leq (\Delta(G)-1)(k-1)+1.$$
Moreover, the bound is sharp and can be obtained in
$\mathcal{O}(kn)$ time.
\end{thm}
\begin{pf} We use Algorithm 1 to prove our bound. In our algorithm, let $Y_i$
be the set of edges incident with $v_i$ in $G_i$ for $1\leq i\leq
k$. Note that in line 3 of our algorithm, we can choose the vertex
$v$ in the maximum component of $G_i$ as $v_i$ such that $v$ is
adjacent to some vertex of $\{v_1, v_2, \cdots,v_{i-1}\}$, so the
vertex $v_i$ must exist. When the algorithm terminates, the final
graph is denoted by $G'=G_k$. It is not hard to show that $G'$
contains at least $k$ components. Since the total number edges
deleted during the algorithm is $\sum_{i=1}^{k-1}{|Y_i|}\leq
\Delta(G)+(\Delta(G)-1)(k-2)=(\Delta(G)-1)(k-1)+1$, we have
$\lambda_k(G)\leq (\Delta(G)-1)(k-1)+1.$

For the sharpness of the bound, we just let $G=C_n$. By Lemma
\ref{thm5}, we have $\lambda_k(G)=k=(\Delta(G)-1)(k-1)+1$ in this
case.

It remains to analyze the running time. In line 3 of Algorithm
1, it takes $\mathcal{O}(n)$ time to find the maximum component of
$G_i$ and choose a vertex $v_i$ with degree at most $\Delta(G)-1$ in
this component.  Therefore the total running time is
$\mathcal{O}(kn)$.
\end{pf}

Obviously, by the above argument, the total running time of Algorithm 1 is at most $\mathcal{O}(n^2)$ since $k\leq n$.

\begin{algorithm}[t]
\caption{} 
\hspace*{0.02in} {\bf Input:} 
A connected graph $G$ with order $n$ and maximum degree
$\Delta(G)$.\\
\hspace*{0.02in} {\bf Output:} 
A subgraph $G'$ of $G$ with at least $k$ components.
\begin{algorithmic}[1]
\State Choose any vertex $v$ as $v_1$ in $G_1=G$, set $G_2=G_1-Y_1$. 
\For{$2\leq i\leq k-1$} 
¡¡¡¡\State Choose a vertex $v_i$ with degree at most $\Delta(G)-1$
in the maximum component of $G_i$; $G_{i+1}=G_i-Y_i$; $i:=i+1$.
\EndFor
\State \Return $G_k$
\end{algorithmic}
\end{algorithm}

Recall that we proved that $\lambda_k(G)\leq
{\Delta(G)}{\kappa_{k-1}(G)}$ in Theorem \ref{thmc} and
$\lambda_k(G)\leq (\Delta(G)-1)(k-1)+1$ in Theorem \ref{thmd}. For
some graphs, the inequality ${\Delta(G)}{\kappa_{k-1}(G)}<
(\Delta(G)-1)(k-1)+1$ holds. For example, let $G$ be the second
example in the proof of Theorem \ref{thmc}, we have
${\Delta(G)}{\kappa_{k-1}(G)}=k-1<
(k-2)(k-1)+1=(\Delta(G)-1)(k-1)+1$ in this case. For some other
graphs, the inequality ${\Delta(G)}{\kappa_{k-1}(G)}>
(\Delta(G)-1)(k-1)+1$ holds. For example, let $G$ be a cycle with
order $n\geq 2k$, where $k\geq 3$. We have
${\Delta(G)}{\kappa_{k-1}(G)}=2(k-1)> k=(\Delta(G)-1)(k-1)+1$ in
this case. By Theorem \ref{thmd}, we clearly have the following
corollary.

\begin{cor}\label{thm8}
Let $G$ be a connected $r$-regular graph with order $n$. For $2\leq
k\leq n$, we have
$$\lambda_k(G)\leq (r-1)(k-1)+1.$$
Moreover, the bound is sharp.
\end{cor}

Recall that $f(n,k,t)$ is the minimum size of a connected graph $G$
with order $n$ and $\lambda_k(G)=t$, where $2\leq k\leq n$ and
$k-1\leq t\leq {n\choose 2}-{{n-k+1}\choose 2}$. We now prove the
following result.
\begin{thm}\label{thmb}
For a connected graph $G$, we have
$$n-k+t\leq f(n,k,t)\leq {{n-k+1}\choose 2}+t.$$
Moreover, we have
$$ f(n,k,t)= \left\{
\begin{array}{ll}
n-k+t, & if~t\in \{k-1,k\}\\
{{n-k+1}\choose 2}+t, & if~t\in \{{n\choose 2}-{{n-k+1}\choose
2}-1,{n\choose 2}-{{n-k+1}\choose 2}\}.
\end{array}
\right.
$$
\end{thm}
\begin{pf}
Let $G$ be a connected graph of order $n$ with $\lambda_k(G)=t$. Let
$Y_0\subseteq E(G)$ with $|Y_0|=\lambda_k(G)$ such that $G-Y_0$
contains at least $k$ components, say $G_1, G_2, \cdots, G_{\ell}$,
where $\ell\geq k$.

We claim that $\ell= k$. If not, then $\ell\geq k+1$ and there is an
edge $e_0=xy\in Y_0$ with $x\in \cup_{i=1}^k{V(G_i)}$ and $y\in
\cup_{j={k+1}}^{\ell}{V(G_j)}$. Without loss of generality, we
assume that $x\in V(G_k)$ and $y\in V(G_{k+1})$. Let
$Y_1=Y_0\setminus \{e_0\}$. It is not hard to show that $G-Y_1$
contains the following $k$ components: $G_1, G_2, \cdots, G_k\cup
G_{k+1}$. This means that $\lambda_k(G)\leq |Y_1|=
|Y_0|-1=\lambda_k(G)-1$, a contradiction. Hence, we have $\ell= k$
and so there are exactly $k$ components in $G-Y_0$.

For the lower bound, we have $m(G)= \sum_{i=1}^k{m(G_i)}+t\geq
\sum_{i=1}^k{(n(G_i)-1)}+t=n-k+t$. Hence, $f(n,k,t)\geq n-k+t.$

For the upper bound, we have that $m(G)= \sum_{i=1}^k{m(G_i)}+t\leq
{{n-k+1}\choose 2}+t$ by Lemma \ref{thm01}. Hence, $f(n,k,t)\leq
{{n-k+1}\choose 2}+t.$

By Lemma \ref{thm5} and Propositions \ref{thm7}, we have
$$
f(n,k,t)= \left\{
\begin{array}{ll}
n-k+t, & if~t\in \{k-1,k\}\\
{{n-k+1}\choose 2}+t, & if~t\in \{{n\choose 2}-{{n-k+1}\choose
2}-1,{n\choose 2}-{{n-k+1}\choose 2}\}.
\end{array}
\right.
$$
\end{pf}

Note that in Theorem \ref{thmb}, we give an upper bound and a lower
bound for the function $f(n,k,t)$. The upper bound can be attained
when $t\in \{{n\choose 2}-{{n-k+1}\choose 2}-1,{n\choose
2}-{{n-k+1}\choose 2}\}$, and the lower bound can be attained when
$t\in \{k-1,k\}$.


\section{Concluding remarks}

In this paper, we investigate the $k$-edge-connectivity
$\lambda_k(G)$ of a graph $G$ and compute some exact values and sharp bounds
 for $\lambda_k(G)$ in terms of some other graph
parameters, such as $\lambda'_k(G)$ and $\kappa'_{k-1}(G)$, where
$2\leq k\leq n$. Specially, we prove that $k-1\leq \lambda_k(G)\leq
{n\choose 2}-{{n-k+1}\choose 2}$ and characterize the graphs with
$\lambda_k(G)=t$, where $t\in \{k-1,{n\choose 2}-{{n-k+1}\choose
2}-1,{n\choose 2}-{{n-k+1}\choose 2}\}$. Then the following problem
is interesting.

\begin{op}\label{op01}~\\
$(i)$~Determine the graphs with $\lambda_k(G)=t$ for a small integer
$t$, that is, $t$ is close to $k-1$.\\
$(ii)$~Determine the graphs with $\lambda_k(G)=t$ for a large
integer $t$, that is, $t$ is close to ${n\choose 2}-{{n-k+1}\choose
2}$.
\end{op}

Recall that we compute the precise values for the
$k$-edge-connectivity of some graph classes which can be used in the results for general graphs. Products of graphs occur naturally in discrete mathematics as tools in combinatorial constructions, they give rise to important classes
of graphs and deep structural problems, and they also play a key
role in design and analysis of networks \cite{Chapman-Nabi-Abdolyousefi-Mesbahi, Imrich}. Some researchers
have investigated a generalized product of graphs \cite{Bermond} and the connectivity which can model and construct large reliable networks under optimal conditions in the past several decades \cite{Piazza,Piazza-Ringeisen
}. It is also interesting to obtain some sharp
upper bounds for the $k$-edge-connectivity of generalized graph products in
terms of some parameters of original graphs, such as the order and
the minimum degree.




We further study the function $f(n,k,t)$ which is defined as the
minimum size of a connected graph $G$ with order $n$ and
$\lambda_k(G)=t$, and give bounds and some exact values for this
function. Since it is quite difficult to determine the exact values
of $f(n,k,t)$ for a general $t$, it is interesting to try the following problem.

\begin{op}\label{op04}~\\
$(i)$~Determine the exact values of $f(n,k,t)$ for a small integer
$t$, that is, $t$ is close to $k-1$.\\
$(ii)$~Determine the exact values of $f(n,k,t)$ for a large integer
$t$, that is, $t$ is close to ${n\choose 2}-{{n-k+1}\choose 2}$.
\end{op}


\end{document}